# DO THE SIZE EFFECTS EXIST?

A.I. Kuklin[1,2], A.V. Rogachev[1,3], A.Yu. Cherny[1], E.B. Dokukin[1], A.Kh. Islamov[1], Yu.S. Kovalev[1], T.N. Murugova[1], D.V. Soloviev[1,4], O.I. Ivankov[1,4], A.G. Soloviev[1], V.I. Gordeliy[1,2,5]

[1]*Joint Institute for Nuclear Research, Dubna, Moscow reg., Russia, Email: kuklin@nf.jinr.ru, Email: rogachev@nf.jinr.ru, Email: cherny@theor.jinr.ru Email: valentin.gordeliy@ibs.fr*

[2]*Moscow Institute of Physics and Technology, Dolgoprudny , Moscow reg., Russia*

[3]*Skobeltsyn Institute of Nuclear Physics of Lomonosov Moscow State University, Moscow, Russia*

[4]*Taras Shevchenko Kyiv National University, Kyiv, Ukraine*

[5]*IBI-2, Forshungszentrum, Juelich , Germany*

Abstract. In this short paper we review a series of publications, some of which are our own, where various aspects of size effects were examined. By analyzing a series of examples we show that various intensive macroscopic characteristics of nanoobjects exhibit non-trivial size dependencies on the scale of 200 to 40 Å. Drastic variations take place for sizes in the region 50-60 Å for ordinary systems, and 60-200 Å in the case of magnetic systems. We argue that X-ray and neutron scattering gives an excellent metrological support in the domain from 100 Å to 10 Å.

## 1. INTRODUCTION

The rapid development of nanotechnologies provides a new impetus for research of nanodisperse structures and for investigations of related scientific problems [1, 2]. One can expect to discover size effects on the scale of typical sizes of supermolecular structures. The quantitative characteristics of the structure of nanodisperse objects are the basis for understanding their properties and functioning. In recent years attention of researchers have been focused on polymers that are interesting from the fundamental science point of view, and also have important practical applications. For instance, dendrimers are objects of investigation within a novel scientific discipline – nanochemistry [2]. These polymers can be used as nanoreactors for synthesis of metallic nanoparticles [3, 4]. Their prospects for biological and medical applications are also actively discussed [5-8]. A similar situation can be observed in the track membrane research. They already are widely applied in industry, medicine, and science [9-12], however, concrete values of, for instance, poly- and monodispersity of highly oriented tracks are largely debatable. The lack of precise knowledge of their structure curbs development of novel nanodisperse objects, which are more efficient in applications.

The book [13] considers a series of problems related to size dependence effects. The authors of that book claim that each intensive physical quantity has its own characteristic size, where non-trivial size effects become apparent.

Below we review several publications by experimentalists, including some of our own papers, where it is demonstrated that the interval from 200 to 30-40 Å is the range where the size effects are observed.

## 2. SIZE EFFECTS AND PROPERTIES OF SUBSTANCES

Detection of the size effects is a quite complex problem due to the absence of a solid metrological support in the range of 100 to 10 Å. Microscopy methods and scattering methods have their own advantages and drawbacks, which essentially determine their domains of applicability. Table 1 summarizes comparative characteristics of these methods.

The main advantages of microscopy methods are visibility of results and the fact that the results are related to the real space. The advantages of scattering methods are the possibility to investigate bulk properties for any state of matter, as well as the possibility to obtain information on the local structure of particles.

The book [13] describes several size effects. The authors claim that various physical characteristics of microscopic clusters attain values typical for bulk materials once their sizes exceed a certain threshold value. In turn cluster sizes depend on the values of measured characteristics. Quite possibly, in order to isolate the essential features of this phenomenon one has to use dimensionless quantities.

We think that the presence of such geometrical thresholds for the parameter values follow already from the observed effects [13 pp. 131, 92, 84, 167]. Non-trivial variations begin on the scale 200 Å, while the major changes take place at 40 Å. The following physical characteristics were considered: melting temperature, (normalized) Young's modulus, the ratio of elementary cell axis sizes, magnetoresistance. As a rule, there is no sharp boundary for size effects; however, it is possible to estimate the threshold value using two straight lines, which extrapolate the data obtained above 200 Angstrom and below 50Å. The intersection point of these straight lines is the desired threshold value. Table 2 summarizes the data from the book [13] obtained this way.

Table 1
Comparison of the structural methods

| № | Parameter | SANS | SAXS | Microscopy |
|---|---|---|---|---|
| 1 | Type of characteristics | Bulk properties | Bulk properties | Surface properties |
| 2 | Object size range | 10-10000Å | 10-10000Å | 10Å- $10^7$ Å |
| 3 | Type of the observed space | Reciprocal space | Reciprocal space | Real space |
| 4 | Range of view | Volume | Volume | Surface, particle on the surface or in volume |
| 5 | Types of samples | Gas, Liquid, Solids, only with good neutrons contrast | Gas, Liquid, Solids, only with good electron density contrast | Solids are preferable |
| 6 | What can be obtained? | Size and shape of particles in solution or matrix, size distribution, density inside of particle, structure factor, molecular weight, number aggregation, fractal dimensions. Magnetic structures | Size and shape of particles in solution or matrix, size distribution, density inside of particle, structure factor, molecular weight, number aggregation, fractal dimensions. | Size, shape and internal structure of local particle in real space, size distribution from a lot of different measurements |
| 7 | What is directly measured? | Nuclear and magnetic contrast | Electron density contrast | Electron density contrast, light scattering contrast, atomic density |
| 8 | Disadvantages | Low sensitivity and resolution, difficult for interpretations. | Low sensitivity and resolution, difficult for interpretations | Local point of view, atomic resolution on surface only, special sample preparation. |

The data presented in Table 2 imply that the average cluster size where drastic variations of physical characteristics are observed is around 60Å. It seems quite likely, that this phenomenon is related to variations of the volume-to-surface ratio and the number of the molecules or atoms in an object. Various size effects are also possible when molecules are placed in a cavity, which dimensions are comparable with the molecule size.

Table 2
Cluster (grain) threshold sizes for different characteristics

| | Physical characteristics | Threshold size (the cusp) Å | References |
|---|---|---|---|
| 1 | The ratio of the axis sizes of elementary cell | 50-60 | [13] p. 84 |
| 2 | Melting point of gold | 40-50 | [13] p. 92 |
| 3 | Young’s modulus (normalized) | 60-70 | [13] p. 131 |
| 4 | Magnetoresistance | 40-80 | [13] p. 167 |

In order to obtain the concrete values of threshold sizes where novel properties begin to emerge, one has to perform the experimental studies that give us the properties as well the size of an object. Such approach was used, for instance, in experiments performed on small-angle scattering device SANS in Dubna. It is well known that water is an important constituent part of majority of biological objects and solutions. The volumes filled by water are often comparable to those where the presence of size effects can be expected. In these cases straightforward applications of known macroscopic values of intensive parameters for description of this sort of water impregnations do not produce adequate results. For this purpose the property of surface-active materials (surfactants) to form reverse micelles in hydrophobic solvents, such as, for instance, $C_6H_6$ and $CCl_4$, was used in experiments described in the papers [14, 15]. For example, at the critical micelle-formation concentration and above, AOT surfactant (Sodium 1,4-Bis(2-ethylhexyl) Sulfosuccinate) forms reversed globular micelles in benzene and decan. Adding water to these substances leads to appearance of reversed globular micelles with central aqueous core. The radius of these micelles is given by $R = A + K(C_{aq}/C_s)$, where A, K – are coefficients, $C_{aq}$ and $C_s$ are the water and surfactant concentrations, respectively. Thus, by changing the concentration ratio $C_{aq}/C_s$ in the system, one can conduct a controlled preparation of stable water droplets with the required radius in the range ~1-100 nm. The volume of small-radius droplets obtained this way significantly exceeds 30 Å$^3$, the well-known volume of the water molecule. As the size of the aqueous core increases, a sharp reduction of the volume takes place, and for the core radii within the bounds 20 Å $< R_h <$ 30 Å the macroscopic value of 30 Å$^3$ is attained. Within the limits of measurement errors this value does not change anymore with a subsequent growth of the aqueous core size. Unfortunately, an extremely important for this kind of measurements, the question of polydispersity was not discussed by the authors of Refs. [14, 15]. Nevertheless, the obtained value of 40-60 Angstrom (for the threshold droplet diameter) is in a good agreement with the values reported by other authors, which experimental results were presented above.

The emergence of a new polymer class – the dendrimers – has led to appearance of a new possibility for creating monodispersed objects. As was shown in Refs. [16,17], the solvent could penetrate the dendrimer molecules. A mathematical model describing the scattering curves for higher generation dendrimers was proposed in Ref. [18].

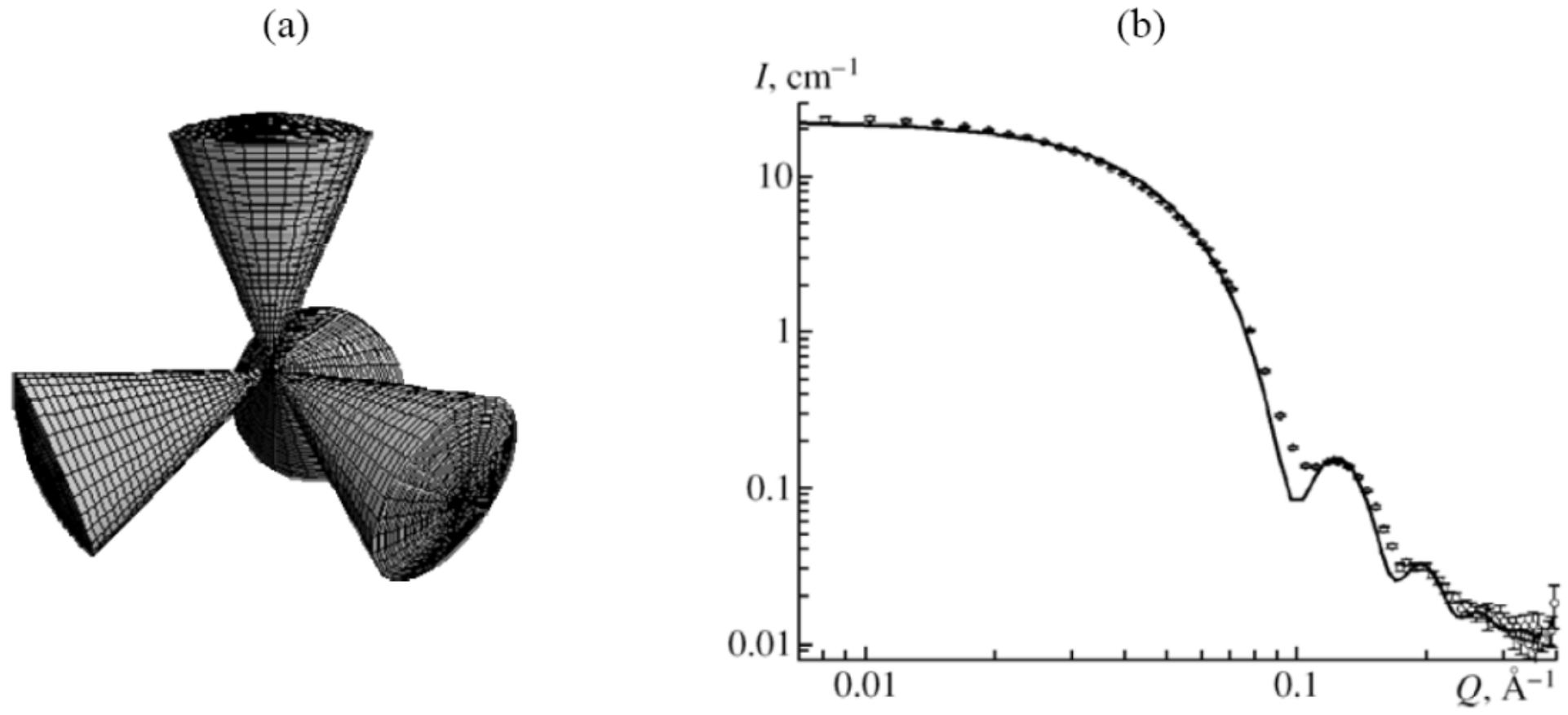

**Figure 1 - (a) A pictorial representation of the spherical sector model. (b) The experimental scattering curve from polycarbosilane dendrimers of the 9th generation (circles) and its approximation with the spherical sectors model (solid line) [18].**

The proposed model describes experimental results qualitatively and suggests that the inner sphere of dendrimers is permeable for the solvent, the total volume of hollows in the model is 18%, and the solvent density in hollows differs from that beyond (bulk) a dendrimer by a factor of 2. This conclusion was made in the framework of the proposed model, and it is related to the core density of dendrimers, and therefore, essentially, to the scattering ability of these macromolecules. The size of the dendrimer molecule where the effect is observed is 50-60 Angstrom (the molecule is anisometric).

The size-effect characteristics found for magnetic properties are slightly different. Figure 2 shows the data from Ref. [19]. The small-angle scattering curves can be described in the framework of the Ornstein-Zernike model and the fractal dimension. The length of the magnetic correlations is about 50-60 Å. The results presented in this paper show that the scattering curve variations taking place as the concentration changes from 41 to 46% and higher lead to a modification of magnetic properties. The estimate for dimensions where the fractal dependence appears is 120-130 Å. This value is very close to the size of the magnetic clasters of Zinc ferrite, for instance. Approximately the same values were presented in Ref. [13], see page 159. The data describing the behavior of Zinc ferrite saturation magnetization as a function of particle size were also presented. A comparison of the data presented in Table 2 and those reported in [19] and [13] indicate that size effects in magnetic materials take place on a different spatial scale.

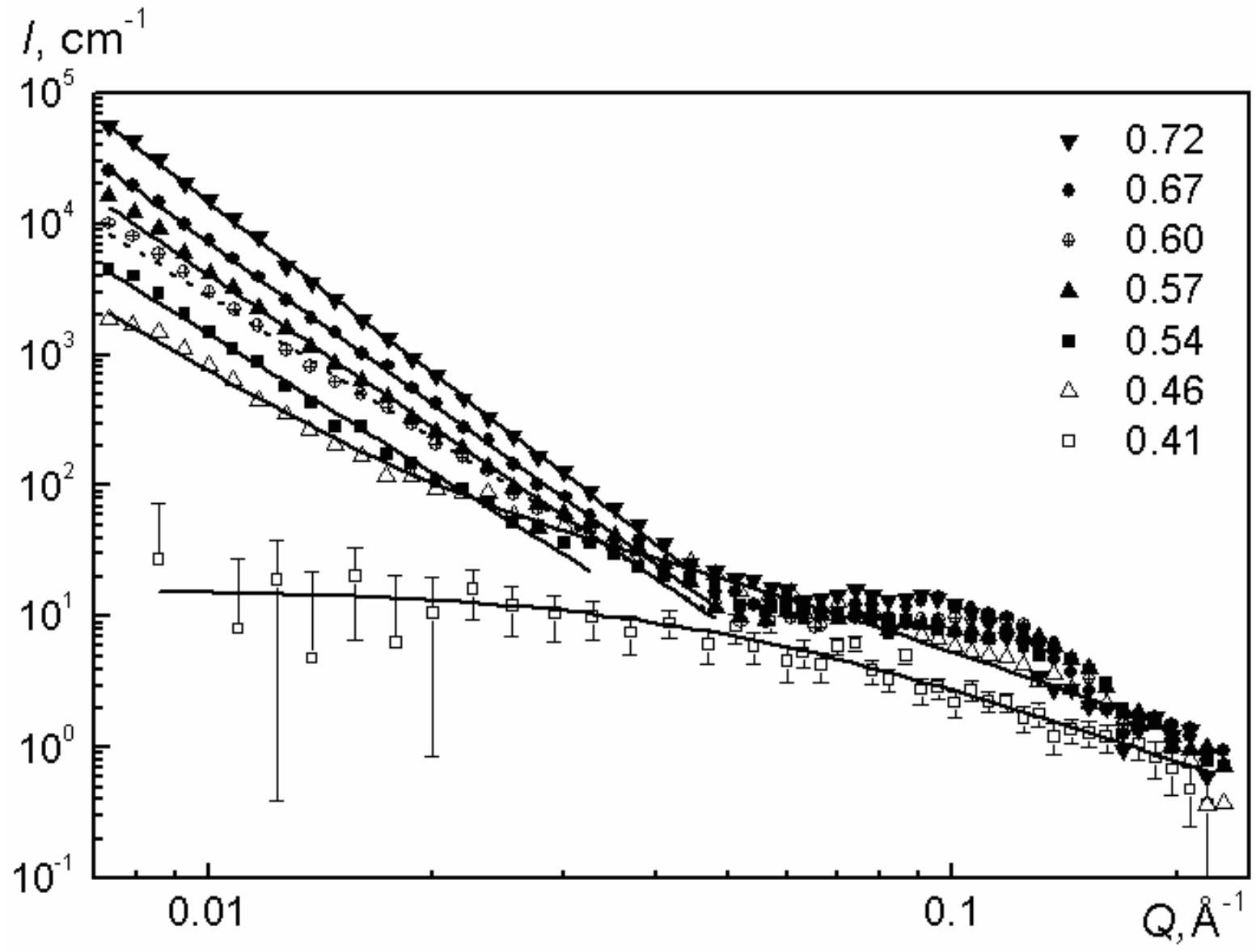

**Figure 2 - Small-angle neutron scattering curves for the nanocomposite $(Co)_x(SiO2)_{1-x}$ – films with Co concentrations $x$ = 41, 46, 54, 57, 60, 67, and 72%)**

## 3. NANOLITHOGRAPHY AND THE NANOMATERIAL REQUIREMENTS

With the development of nanotechnologies the traditional domains of the track membranes applications (filtering, ultrafiltering, single lipid membranes (vesicles) preparation and others) can be significantly expanded. For instance, application of track membranes and the "pinhole camera" technique allowed one to obtain nanoobjects by sputtering, see [11]. The authors not only proposed a nanotechnology project – nanolithography, but they have completed it as well.

The resolution of 50 nm has been achieved in their experiments. The Cr atom objects have clear-cut shapes and dimensions around 70 nm. In this case the membrane polydispersity determines the polydispersity of the obtained objects, their dimensions are determined by pore sizes, while the channel roughness affects the image boundaries (replicas). Thus, the application of TM in nanotechnologies depends essentially on the film quality. As will be shown below, small-angle scattering allows one to measure with a high precision poly- and monodispersity, as well as the roughness of the membrane surface, that is, the properties which determine the quality of materials for industrial applications. However, nanolithography processes require membranes with good monodispersity. The experiments on the quantitative monodispersity degree measurements were reported in the papers [20-22]. A novel methodology was developed there, and materials with extremely high monodispersity degree of track pores were found. It is this kind of track membranes that are required for nanotechnology applications.

## 4. CONCLUSION

This short review paper shows that one can consider the values of 40-60 Å as the size effect thresholds for various physical characteristics of ordinary materials, while for magnetic properties this value is more than two times higher. High monodispersity of the objects used in nanotechnologies often occurs to be a key requirement. Unfortunately, accurate values of the structural characteristics of the objects with nano sizes are not so trivial to obtain. As it has been demonstrated small-angle X-ray and neutron scattering gives us an excellent metrological support in the domain from 100 Å to 10 Å. An important advantage of this method is that it allows one to obtain not only integral structural parameters but the important data about the internal structure of the nanoobjects as well. The latter is necessary to understand unusual properties of these nanoparticles and develop new nanomaterials with the required properties.